\documentclass[12pt]{scrartcl}
\usepackage{amssymb,newlfont}
\def\be{\begin{equation}}
\def\ee{\end{equation}}
\def\bc{\begin{center}}
\def\ec{\end{center}}
\def\bq{\begin{quote}}
\def\eq{\end{quote}}
\def\ket#1{ | #1 \rangle }
\def\bra#1{ \langle #1 | }
\def\braket#1#2{ \langle #1 | #2 \rangle }

\begin{document}
\sloppy

\thispagestyle{empty}

\noindent{\emph{Foundations of Physics. Special issue: Festschrift to Peter Mittelstaedt's 80th birthday}}

\vspace*{9mm}

\bc

\vspace*{5mm}

\renewcommand{\thefootnote}{\fnsymbol{footnote}}

{\Large\sf \textbf{Why Quantum Theory is Possibly Wrong}}

\setcounter{footnote}{0}

\vspace*{12mm}

{\large Holger Lyre}%
\footnote{Philosophy Department,
University of Magdeburg, Germany,
Email: lyre@ovgu.de}

\setcounter{footnote}{0}

\vspace*{11mm}

{\large May 2010}

\ec

\vspace*{8mm}

\bq
\renewcommand{\baselinestretch}{1}\small
{\noindent\textbf{Abstract.}
Quantum theory is a tremendously successful physical theory,
but nevertheless suffers from two serious problems:
the measurement problem and the problem of interpretational underdetermination.
The latter, however, is largely overlooked as a genuine problem of its own.
Both problems concern the doctrine of realism, but pull, quite curiously,
into opposite directions. The measurement problem can be captured such
that due to scientific realism about quantum theory common sense anti-realism follows,
while theory underdetermination usually counts as an argument against scientific realism.
I will also consider the more refined distinctions of ontic and epistemic realism
and demonstrate that quantum theory in its most viable interpretations
conflicts with at least one of the various realism claims.
A way out of the conundrum is to come to the bold conclusion that quantum theory is,
possibly, wrong (in the realist sense).
}
\eq

\vspace{8mm}



\section{Introduction}

Quantum theory (QT) is presumably the most successful theory in the history of physics ever.
It provides the broad theoretical framework for constructive model theories
like quantum electrodynamics or solid state quantum mechanics,
and its various theoretical predictions are as impressive as for instance
the precise calculation of the anomalous magnetic dipole moment
of the electron within an accuracy of $10^{-8}$.
It's perhaps even more instructive to illustrate the success of QT
by pointing out that about one third of the gross national product
of the US is based directly or indirectly on developments of QT.

Nevertheless, QT suffers from two serious problems:
the quantum measurement problem and the problem of interpretational underdetermination,
and both problems concern the philosophical doctrine of realism.
Curiously, and as far as the issue of realism is concerned, the two problems
pull into opposite directions. The measurement problem can be captured such
that if we are scientific realists about QT common sense anti-realism follows,
while theory underdetermination usually counts as an argument against scientific realism.
This reflects more than just a superficial philosophical tension,
it reflects the deep conceptual problems of QT. In fact, the problems are deep enough
to come to the conclusion that quantum theory is possibly wrong.
And this is the overall thesis I will argue for in the paper.

The paper will be organized as follows.
In the second section I will present the quantum measurement problem
and emphasize the points I consider to be important.
The third section is devoted to the general issue of theory underdetermination
and its particular relevance and application to the interpretational debate of QT.
In section four the doctrine of realism will be deployed in its various relevant
distinctions of common sense realism and scientific realism as well as the ontic/epistemic
divide (section 5). I will balance and discuss the six most prominent interpretations
of QT with regard to the different realism variants.
It turns out that QT in any of the considered viable interpretations
conflicts with at least one of the realism variants.
In the final section I will ask what's so special about QT.
The point I want to make is that it is the only theory in science which
leads, according to the measurement problem, to a serious attack on common sense realism
and which at the same time provides the most catchy case study
for the otherwise debatable issue of theory underdetermination.
The final conclusion that quantum theory is possibly wrong follows as
a natural but nevertheless astonishing consequence from our foregoing discussion.


\section{The Quantum Measurement Problem}

The quantum measurement problem (QMP) is by far the most intricate sting in the quantum
business. Loosely speaking, QMP arises from that fact that there is no unitary transition
\be
\ket{\phi} = \sum_{i} \alpha_{i} \ket{\phi_i}  \ \nrightarrow \ \ket{\phi_k}
\ee
with probability $p_i=|\alpha_{i}|^2$ (Born's rule).
Quantum states are generally construed as superpositions (lhs),
while measurement outcomes appear to be definite results (rhs).

Some of the most comprehensive analyses of the problem of measurement in QT
can be found in the lifework of Peter Mittelstaedt (1963, 1998),
the following exposition is very much inspired by his work.
Consider a measurement apparatus $\cal{A}$ and object system $\cal{S}$
with corresponding states $\ket{a}$ and $\ket{s}$.
To perform a measurement, the systems $\cal{A}$ and $\cal{S}$ must be coupled.
Formally, one considers the compound states $\ket{\phi}$ in the
tensor product $\cal{H}_\cal{S} \otimes \cal{H}_\cal{A}$
of the Hilbert spaces $\cal{H}_\cal{S}$ and $\cal{H}_\cal{A}$.
The physical coupling itself is represented by the dynamics
\be
\ket{ \phi'} = e^{ i \hat H t  } \ \ket{ \phi }
\ee
of the measurement interaction ${\hat H}$.
It follows from linearity that for a general initial state of the measured system
we get entangled states
\be
\ket{\phi'} = \sum_{i,k} c_{ik} \ket{a'_i \otimes s'_k }.
\ee
The Schmidt decomposition guarantees that there exists a representation such that
$\ket{\phi'} = \sum_i \sqrt{p_i} \ket{a'_i} \ket{s'_i}$ with $p_i = |c_{ii}|^2$,
nevertheless, $\ket{\phi'}$ is a pure state of the compound system $\cal{A}+\cal{S}$.

Here, the measurement problem arises.
The crucial point is that, after the measurement coupling, systems $\cal{A}$ and $\cal{S}$
are no longer independent, and that therefore the states of the compound $\ket{\phi}$
are not factorizable into the single system states $\ket{a}$ and $\ket{s}$.
This is the root of problem, since we expect from any measurement and measuring apparatus
to yield independent, definite pointer states.
It is a precondition of any measurement that the following premiss holds:
\bq
\textbf{Central measurement premiss (CMP):}
The outcome of any measurement process is given by definite pointer states.
\eq
It follows from the above that this premiss can generally not be fulfilled in QT.
This is (one way to spell out) the measurement problem.
\bq
\textbf{The quantum measurement problem (QMP):}
Quantum theory is in conflict with CMP, since it cannot reproduce definite measurement outcomes.
\eq

It has become standard to consider decoherence mechanisms in order to cope
with QMP (cf. Schlosshauer 2007).
The idea is to embed the system $\cal{A}+\cal{S}$ in a more realistic manner
into a bigger environment $\cal{E}$, where the crucial assumption can be made that
the states of $\cal{E}$ are more or less uncorrelated
\be
\label{decohere}
\braket{e_i}{e_k} \approx \delta_{ik}.
\ee
The total state of the compound $\cal{A}+\cal{S}+\cal{E}$ may be written as
$\ket{\Phi} = \sum_i \sqrt{p_i} \ket{a_i'} \ket{s_i'} \ket{e_i'}$
or as a density matrix $\rho=\ket{\Phi}\bra{\Phi}$.
Under the decoherence assumption (\ref{decohere}) the state of the
subsystem $\cal{A}+\cal{S}$ can be written as the reduced density matrix
\be
\label{rhored}
\rho_{red} \approx \sum_i p_i \ket{a_i'}\bra{a_i'} \otimes \ket{s_i'}\bra{s_i'}.
\ee
Prima facie, this looks promsing, since due to decoherence the disturbing interference terms
have (almost) been deleted. It is, however, well known, that the reduced density matrix $\rho_{red}$
cannot be distinguished from a statistical ensemble of states $\ket{a_i'} \ket{s_i'}$
\emph{for all practical purposes}!
As d'Espagnat (1965) has dubbed it, $\rho_{red}$ is an \emph{improper mixture},
it only \emph{appears as if} a certain measurement result has been achieved.

Hence, decoherence alone cannot solve the measurement problem.
It is instructive to decompose QMP into a two-fold problem:
\begin{enumerate}
\item Singling out (the pointer basis as) a preferred basis system
\item Transformation of a pure into a non-pure state.
\end{enumerate}
While decoherence offers an explanation to problem (1) and thereby nicely explains
why the actual world appears as classical, decoherence has no resources to solve (2).
As John Bell (1990) once put:
\bq
\emph{``The idea of elimination of coherence, in one way or another,
implies the replacement of `and' by `or', is a very common one among
the solvers of the `measurement problem'. It has always puzzled me.''}
\eq

QMP is therefore fresh and alive. And there are a few ways to express it.
Here are some corollaries:
\begin{itemize}
\item There exist no unitary mappings from pure states to mixed states.
\item Quantum measurements lead to improper mixtures only.
\item Quantum state probabilities do not allow for an ignorance interpretation, they are ontic probabilities.
\item Quantum theory doesn't provide its own measurement theory.
\end{itemize}

It should have become clear from the thus exposed nature of QMP
that QT is truly unique among all physical, or even among all scientific
theories in the sense that no other theory is plagued by such an intricate problem.


\section{Quantum Theory Underdetermination}
\label{QTUD}

Scientific theories give us pictures of the world -- pictures of the world beyond mere
collections of sense data and observations. Such ontological pictures are given to us
if we provide the theoretical formalism with an interpretation.
While this is a common theme for any scientific theory, and even more so, of course,
for mathematically formalized theories, quantum theory is unique in this respect, too.
No other scientific theory has ever been plagued in the same sense and to the same extent
by the problem of giving an appropriate interpretation of the formalism.
80 years of QT have provided us with a large variety of differing interpretations.
Here's a rough and ready list of just a few common ones:
instrumentalism, statistical (ensemble) interpretations, Copenhagen interpretation,
consciousness-caused collapse interpretations (\`a la Wigner),
many worlds (\`a la Everett), many minds, many histories, consistent histories,
Bohmian mechanics, spontaneous collapse theories (\`a la Ghirardi-Rimini-Weber),
transactional interpretation, relational quantum mechanics, modal interpretations etc.

For the following let us pick out six interpretations out of the whole variety,
but basically \emph{pars pro toto}. They are
\begin{enumerate}
\item Instrumentalism
\item Copenhagen interpretation
\item Many worlds
\item Bohmian interpretation
\item Consciousness-caused collapse interpretations
\item Spontaneous collapse theories (GRW)
\end{enumerate}
These six interpretations give us drastically heterogeneous ontological pictures of the world,
but are, at the same time, empirically equivalent in the sense that they satisfy the same
corpus of observational data. At least, we can recast them in such a way that this claim holds.
In a slogan: they are empirically equivalent, but ontologically different.

In a sense GRW sticks out.
GRW-like approaches do in fact change the mathematical core of the formalism by adding
a new additional piece, the collapse mechanism, to it. Nevertheless GRW-like approaches
can in principle be adjusted in such a way that they fit the same data as interpretations 1-5.
This at least works up to a point far beyond today's measuring accuracies.
In this sense all six interpretations provide cases of empirically equivalent,
but ontologically different variants of QT, and as such intriguing cases of what
philosophers of science call theory underdetermination by empirical evidence.

In short, the thesis of theory underdetermination (TUD) says the following:
\bq
\textbf{TUD-Thesis:}
For any theory T and any body of observation O there exists another theory T',
such that T and T' are empirically equivalent (but ontologically different).
\eq

The main intuition behind TUD is that theory exceeds observation ($T>O$).
Theories are far more than mere collections of data or listings of outcomes of experiments,
theories introduce theoretical terms and lawlike connections between them,
either as logical connections or as empirically grounded regularities.
It is the slack between T and O which, in principle at least,
allows for a multitude of ways to fit the data with theory.
This basic intuition behind TUD is beautifully captured in Quine's words in his
1975 paper ``On empirically equivalent systems of the world'':
\bq
\emph{''If all observable events can be accounted for in one comprehensive scientific theory--one system
of the world...--, then we may expect that they can all be accounted for equally in another,
conflicting system of the world. We may expect this because of how scientists work.
For they do not resist with mere inductive generalizations of their observations:
mere extrapolations to observable events from similar observed events.
Scientists invent hypotheses that talk of things beyond the reach of observation.
The hypotheses are related to observation only by a kind of one-way implication;
namely, the events we observe are what a belief in the hypotheses would have led us to expect.
These observable consequences of the hypotheses do not, conversely, imply the hypotheses.
Surely there are alternative hypothetical substructures that would surface in the same observable ways.
Such is the doctrine that natural science is empirically under-determined ... by all observable events.''
}\eq

In the debate about scientific realism, TUD is usually considered as one of the strongest objections
to the realist position (besides the equally infamous pessimistic meta-induction).
It is also important to notice that TUD is a particularly strong claim.
This becomes clear if we compare it with neighboring, though not equivalent claims.
For instance, TUD should not be confused with Duhemian holism -- the claim that
there is no \emph{experimentum crucis}, that no scientific hypothesis can be tested in isolation,
but only theories as a whole. According to such confirmational holism it is possible
to adhere to any thesis in the face of adverse observations by revising other theses.
Only whole theories are subject to confirmation. Surely there is only a small gap to TUD,
since we may very well generate rivaling theories by readjusting the total system of hypotheses.
According to TUD, however, even the total system cannot be confirmed,
but is underdetermined by \emph{all} possible observations.

To emphasize that TUD speaks about underdetermination by \emph{all} possible observations
shows the difference to the induction problem, sometimes dubbed as Humean underdetermination.
The induction problem is induced by underdetermination of theory by past evidence,
while TUD considers underdetermination even in the case of all possible (past and future) observations.

As a strong claim, TUD is by far not uncontroversial. The most pressing problem with TUD
as a convincing objection to scientific realism is the perplexing fact that there doesn't
seem to exist that many convincing cases. In fact, given the generality of TUD as stated
in the philosophy literature one would assume that practising scientists suffer seriously from it.
One would assume that science in practice is notoriously stymied by the appearance
of rivaling theories, that scientists always have to cope with conflicting theoretical models.
But that doesn't seem to be the case -- certainly not as far as physics is concerned.
By and large, science seems to be conservative and calm.
This is what we might call the problem of missing examples (Lyre 2009).

As we've seen from the above, however, the interpretational debate in quantum theory is a counterexample.
The plethora of rivaling quantum interpretations provides us with the perhaps only convincing TUD case in physics.
Call this QTUD: the underdetermination of quantum theory by empirical evidence as displayed in the
multitude of rivaling interpretations.


\section{Common sense realism and scientific realism}
\label{CSR-SciR}

TUD is a threat to scientific realism. Scientific realism, in turn, is a particular realist doctrine.
It is the doctrine that the theoretical terms in our best and mature scientific theories refer.
It is therefore a claim about the unobservable. By way of contrast, common sense realism is the claim
that the things in our directly observable, common sensical world around us exist.
Common sense realism is about tables and chairs,
while scientific realism concerns theoretical entities like electrons or black holes.
Let us, for the sake of clarity, capture the two doctrines in brief:
\bq
\textbf{Common sense realism (CSR):}
There exists a world of everyday (``mesoscopic'') and concrete, particular entities
such as tables, chairs, animals, humans, trees, mountains, the sky etc.
\eq
\bq
\textbf{Scientific realism (SciR):}
The theoretical terms in our best and mature scientific theories refer.
\eq

Obviously, in discussions about realism one must be careful about the divide
between common sense realism and scientific realism, since it simply makes
an important difference whether we talk about tables and chairs or theoretical entities.
And what is more: scientific anti-realists may very well be (and usually are!) common sense realists.
We may even say that science in general is actually committed to common sense realism.
The reason for this commitment simply lies in the fact that scientific experimentation and
theorizing presupposes belief in the reality of our everyday, mesoscopic world.
It is more than obvious that disbelief in the reality of our laboratories, measuring devices
and pointer states doesn't even get the scientific enterprise off the ground.
And this, in turn, means that scientific realism presupposes or includes common sense realism.
Let us capture this in the following way:
\be
\label{SciR-supset-CSR}
SciR \supset CSR
\ee

In the light of the quantum measurement problem, however, the existence of definite states
and hence the independent and objective existence of objects represented by such states
is undermined. There is, furthermore, no restriction about the realm of objects described
by quantum theory, QT is generally thought to be universally applicable.
Thus, QMP is a threat to realism \emph{in toto}, it in fact threatens CSR:
\be
\label{QMP-supset-negCSR}
QMP \supset \neg CSR
\ee
And this is quite remarkable, since QT really turns out to be special amongst scientific theories
in this respect: it is the only scientific theory that threatens CSR!
We may consider this as another, more informal way to spell out the quantum measurement problem.


\section{Realism and quantum interpretations}

The various quantum interpretations offer various ways to cope with the measurement problem.
For instance: Bohm -- by assuming hidden variables, many worlds -- by assuming many worlds,
and GRW -- by assuming additional collapses.
But here we do not want to ask whether the QT variants succeed in solving QMP,
but rather ask: do these interpretations stick with CSR?
More generally: how do the various QT-interpretations stick with either common sense realism
or scientific realism about QT, SciR(QT)?
The following table might give a first, rough overview:
\bc
\begin{tabular}{l|c|c}
                  & CSR  & SciR(QT) \\
  \hline
  Instrumentalism & +    & - \\
  Copenhagen      & -    & + \\
  Bohm            & +    & + \\
  Many worlds     & -    & + \\
  Wigner          & -    & + \\
  GRW             & +    & + \\
  \hline
\end{tabular}
\ec

The first line is straightforward: instrumentalists are scientific anti-realists,
but are at the same time (or may at least be) common sense realists.
Proponents of the Copenhagen interpretation take QT seriously, not only as a tool,
but as telling us something about the world. They do, however, reserve a special
role for observers or measuring devices or conceive a classical/quantum divide.
In short: SciR(QT) yes, CSR no.
The same is true for many world proponents, who assume,
quite contrary to common sense, a branching of worlds into many copies,
and Wignerians, who refer to a mind-body-dualism.

While 4 of 6 interpretations violate some form of realism, Bohm and GRW do not.
But we must be more careful in our usage of the term realism and distinguish at least
between ontic and epistemic variants of realism. For the purpose of our discussion
this distinction is particulary relevant in the case of scientific realism.
\bq
\textbf{Ontic SciR:}
The theoretical entities in our best and mature scientific theories exist independently
(from our epistemic and linguistic capacities).
\eq
\bq
\textbf{Epistemic SciR:}
Science conveys true knowledge about theoretical entities.
\eq

Under these refined definitions the preceding table must be enlarged in the following way:
\bc
\begin{tabular}{l|c|c|c}
                  & CSR & OSciR(QT) & ESciR(QT) \\
  \hline
  Instrumentalism & +   & -         & - \\
  Copenhagen      & -   & +         & - \\
  Bohm            & +   & +         & - \\
  Many worlds     & -   & +         & + \\
  Wigner          & -   & +         & - \\
  GRW             & +   & +         & + \\
  \hline
\end{tabular}
\ec

Under the refined definitions the Bohmian approach also violates a form of realism, ESciR,
since Bohmians build their approach on the assumption of epistemically hidden variables.
It seems that, in the end, only GRW conforms to all realistic assumptions.
However, we should not forget the very special nature of GRW-like approaches as already
indicated in section \ref{QTUD}.
GRW adds an explicit new mechanism to QT, the collapse mechanism. Without going into details,
such a mechanism typically depends on new parameters specifying the localization accuracy and
the mean frequency of the stochastic spontaneous collapses.
The collapse parameters are put in by hand such that the predictions of the model
conform to the experimental outcomes.
They are \emph{ad hoc}, but this may be considered an aesthetic stain of GRW only.
The deeper problem is that it seems in principle always possible to bring GRW approaches
in accordance with observation by suitable readjustments of the collapse parameters.
GRW provides therefore an all-too obvious case of Duhemian holism:
the readjustment procedure can be used to make GRW auto-immune to any falsification
(at least up to a point beyond today's measuring accuracy, as already mentioned in section \ref{QTUD}).

Obviously, if true, this is no good science. And if this is the price to pay for realism
then realism becomes a mute point. Let us therefore concentrate on interpretations 1 to 5 only.
Under this restriction, the upshot of this section is that quantum theory in its most viable
interpretations conflicts with at least one of the various realism claims.
Besides GRW with its contentious setting, there seems to be no hope to reconcile QT
with a full-blown realism: CSR + OSciR(QT) + ESciR(QT).


\section{What's so special about quantum theory?}

Section \ref{CSR-SciR} ended with the result (\ref{QMP-supset-negCSR}) that QMP threatens CSR,
and we have considered this as an informal way to spell out QMP.
We also saw that QT is special among scientific theories since it is the only scientific theory
that threatens CSR (at least to this extent).
On closer inspection, however, this is a weird result, since in order
to believe in QT we have to be scientific realists first!
Our result from section \ref{CSR-SciR} says that scientific realism
about quantum theory, SciR(QT), ``infects'' common sense realism
\be
\label{SciR-supset-negCSR}
SciR(QT) \supset QMP \supset \neg CSR.
\ee
On the other hand, from (\ref{SciR-supset-CSR}), SciR in general already includes CSR.
Thus, in a sense, (\ref{SciR-supset-CSR}) and (\ref{SciR-supset-negCSR}) 'contradict' each other.
This may not be a strict logical contradiction, but there is certainly something strange going on here.

So let's try over. QT is special. It is special because it suffers not only from QMP
but from \emph{two} serious problems: \textbf{QMP} and \textbf{QTUD}.

Some will argue that both problems are connected, and that's perhaps true.
But if so, how are they connected?
Is the measurement problem an aftereffect of QTUD? This seems largely implausible.
Theory underdetermination, the general fact that there exist ontologically different
but empirically equivalent rivals, does not lead in any logical or conceptual sense
to a particular, concrete problem like QMP. So $QTUD \supset QMP$ is quite unlikely.
But the reverse,
\be
\label{QMP-supset-QTUD}
QMP \supset QTUD,
\ee
is by no means inconceivable.

Notice first that, even if we assume the two problems QMP and QTUD to be connected,
they seem to pull into opposite directions.
We saw that the measurement problem can be captured such that, according to (\ref{SciR-supset-negCSR}),
if we are scientific realists about QT, common sense anti-realism follows,
while theory underdetermination usually counts as an argument against scientific realism,
hence $TUD \supset \neg SciR$.
In the case of quantum theory, however, scientific realism amounts to taking
the measurement problem seriously: $SciR(QT) \supset QMP$,
compare again (\ref{SciR-supset-negCSR}). Combining this, we get the curious chain
\be
QMP \supset QTUD \supset \neg SciR(Q) \supset \neg QMP.
\ee
Again, this looks weird.
And even if our rather informal notation ''$\supset$'' doesn't mean strict logical implication,
we must somehow break the chain of arguments at a certain point to avoid the contradictory flavor
in all this. Is there a possible way out of the conundrum?

Let's go back to (\ref{QMP-supset-QTUD}), which we used as a conceivable assumption in our argument.
Why is it conceivable? QMP, or so one might argue, points to a deep-seated
inconsistency or fallacy within quantum theory. And if that's true, QT must basically
be considered an incomplete and scientifically underdeveloped framework. It is therefore
by no means astonishing that we can cook up so many rivaling interpretations for QT,
in the light of this consideration they simply appear as an outspring of
incomplete scientific knowledge.

However, this means that QTUD is presumably not the genuine case of TUD we mistook it for.
We should not expect incomplete theoretical frameworks completely to be interpretable.
We must, on the contrary, expect such frameworks to allow for an open-ended multitude
of conflicting interpretations. And even if the inference (\ref{QMP-supset-QTUD}) cannot
be made strongly, QTUD is certainly ''triggered'' by QMP to a considerable extent.

This line of reasoning also offers a potential answer to the question, why QTUD appears to be
the only half-way convincing case of TUD in physics as pointed out in \ref{QTUD}:
TUD should be omnipresent, but there is a problem of missing TUD examples instead.
The reason might simply be that it takes a conceptual problem in science as deep as QMP
to ``trigger'' rivaling frameworks. Without such a problem they practically never appear.

Of course we can understand this either epistemically or ontologically.
Under an epistemic reading the claim only is that the rivaling frameworks do not come to light,
but that they nevertheless exist. Under an ontological reading they only exist because of
the previously existing conceptual difficulties. The latter option is perhaps more plausible.
At least for the practice of science the idea of Quinean TUD seems to be overdrawn.
Practical cases of TUD, which are rare events anyhow, should not be considered as
genuine TUD cases in the strong Quinean sense,
but rather as indicators of incomplete scientific knowledge (cf. Lyre and Eynck, 2003).
Under this supposition QTUD isn't genuine as well, but an artefact of conceptual incompleteness.
And this means that we must seriously envisage the conclusion that \emph{quantum theory is possibly wrong}.

Let me finally remark that this conclusion is of course as vague as many of the assumptions
made along the way. To say that QT is wrong is meant here in the naive correspondence sense:
the theoretical terms of a wrong theory simply do not refer. To say that QT is \emph{possibly} wrong
indicates that I'm fully aware of the contentious nature of some of my assumptions.
Nevertheless, I think it is useful to throw light on old questions, the conundrum of QMP,
from a somewhat different and largely overlooked angle, the question of QTUD.
But surely, to claim that QT is possibly wrong is a bold claim,
since QT is so tremendously successful.
The debate about quantum realism is still open.



\section*{Acknowledgement}

I dedicate this paper to Peter Mittelstaedt -- a constant source
of scientific inspiration and personal warmth to me over more than 15 years.
For this I wish to offer my heartfelt thanks to him.
I also like to thank Paul Busch for helpful comments
on an earlier version of the paper.

\section*{References}

\begin{description}

\item[]
Bell, J.~S. (1990): Against `Measurement'. Physics World 8:33-40.

\item[]
Espagnat, B. d. (1965): Conceptions de la physique contemporaine. Hermann, Paris.

\item[]
Lyre, H. and Eynck, T.~O. (2003):
Curve it, Gauge it or Leave it? Practical Underdetermination in Gravitational Theories.
\emph{Journal for General Philosophy of Science} 34(2):277--303.

\item[]
Lyre, H. (2009):
Is Structural Underdetermination Possible?
Synthese (Online first, DOI 10.1007/s11229-009-9603-z)

\item[]
Mittelstaedt, P. (1963).
\emph{Philosophische Probleme der modernen Physik}.
B.I.-Wissenschaftsverlag, Mannheim.

\item[]
Mittelstaedt, P. (1998).
\emph{The Interpretation of Quantum Mechanics and the Measurement Process}.
Cambridge University Press, Cambridge.

\item[]
Schlosshauer, M. (2007): Decoherence and the Quantum-to-Classical Transition Springer, Berlin, 2007.

\item[]
Quine, W.~V. (1975).
On Empirically Equivalent Systems of the World.
\emph{Erkenntnis} 9:313--328.

\end{description}


\end{document}